\documentclass[preprint,12pt]{revtex4-1}
\usepackage{amssymb}
\usepackage{amsmath}
\usepackage{dsfont}
\usepackage[T1]{fontenc}
\usepackage{arydshln}

\usepackage{graphicx}
\usepackage{dcolumn}

\usepackage[utf8]{inputenc}
\usepackage{hyperref}

\newcommand{\bq}{\begin{eqnarray}}
\newcommand{\eq}{\end{eqnarray}}
\newcommand{\be}{\begin{equation}}
\newcommand{\ee}{\end{equation}}
\newcommand{\ei}{\mathrm{e}}
\newcommand{\tr}{\operatorname{tr}}

\newcommand{\abs}[1]{\left| #1 \right|}

\newcommand{\Tr}[1]{\mathrm{\text{Tr}}\left[#1\right]}

\begin{document}




\title{Classes of Gaussian States for Squeezing Estimation}


\author{Leonardo A. M. Souza}
\email{leonardoamsouza@ufv.br}
\affiliation{{Federal University of Vi\c{c}osa -- Campus Florestal}, {LMG 818 Km6 S/N},{Florestal},{35690-000}, {Minas Gerais}, {Brazil}}

\begin{abstract}
This study explores a detailed examination of various classes of single- and two-mode Gaussian states as key elements for an estimation process, specifically targeting the evaluation of an unknown squeezing parameter encoded in one mode. To quantify the efficacy of each probe, we employ the concept of Average Quantum Fisher Information (AvQFI) as a robust metric to quantify the optimal performance associated with specific classes of Gaussian states as input. For single-mode probes, we identify pure squeezed single-mode states as the optimal choice and we explore the correlation between Coherence and AvQFI. Also, we show that pure two-mode squeezed states exhibit behavior resembling their single-mode counterparts for estimating the encoded squeezing parameter, and we studied the interplay between entanglement and AvQFI. This paper presents both analytical and numerical results that encompass all the studied classes, offering valuable insights for quantum estimation processes.

\end{abstract}



%


\maketitle

\section{Introduction}\label{intro}

Estimation theory encompasses methods and principles used for making predictions based on limited or incomplete data \cite{kubacek2012foundations}. This theory, often called Metrology, is vital for technological advances in various fields \cite{kubacek2012foundations, kay1993fundamentals}. The core idea of an estimation strategy, be it classical or quantum, is to achieve the the best accuracy in determining a parameter encoded in the system. This often involves using multiple copies of the system for encoding and conducting a precise measurement to approach the true parameter value \cite{kubacek2012foundations, kay1993fundamentals, helstrom1976quantum, holevo2003statistical, ballester2005estimation, paris2009quantum}.

In estimation theory, we can utilize classical or quantum systems for an estimation task and compare their performance, in order to determine if there is a ``quantum advantage'' in estimating unknown parameters. The extensive literature on this subject delves comparisons between classical and quantum systems, shedding light on their individual strengths and responses within the domain of estimation tasks \cite{helstrom1976quantum, holevo2003statistical, paris2009quantum, adesso2009optimal, AdessoIP, Luca, luo2022environmental, predko2020time, assad2020accessible, bradshaw2018ultimate, morelli2021bayesian, ge2020operational, escher2011general, escher2011quantum, maleki2023speed}. It is important to emphasize that within the Estimation Theory, the estimation precision of a parameter is quantified by the Fisher Information (FI), or by its quantum version, the Quantum Fisher Information (QFI) \cite{helstrom1976quantum, holevo2003statistical}. Typically, higher FI or QFI values correspond to better parameter precision. Also, one may indeed obtain a precision incrementation using quantum resources \cite{adesso2009optimal, escher2011general,schnabel2010quantum, demkowicz2013fundamental, pezze2018quantum, giovannetti2011advances, giovannetti2006quantum}.

Quantum Metrology, or quantum estimation theory, the use of quantum systems and measurement apparatus to perform an estimation task, plays a pivotal role in our technological advancements, particularly exemplified by the Laser Interferometer Gravitational-Wave Observatory (LIGO) \cite{abbott2020prospects}: LIGO's profound success in detecting gravitational waves, as predicted by Einstein's theory of relativity, can potentially achieve even greater success by relying on quantum-based precision measurements \cite{yu2020quantum}. Additionally, the utilization of squeezed light in the LIGO experiment can significantly enhance the sensitivity of the detector \cite{aasi2013enhanced, dwyer2022squeezing, grote2013first}. Furthermore, it is fundamental in the design and optimization of quantum-enhanced sensors \cite{niezgoda2021many, marciniak2022optimal,degen2017quantum} and metrology techniques, both theoretical and experimental \cite{pezze2018quantum}.

Within the scope of estimation theory using quantum systems and measurements, one can specify the analysis in discrete and continuous variables (the latter is commonly abbreviated to CV) \cite{adesso2009optimal, pezze2018quantum}. Finally, for continuous variables systems it is possible to subdivide the analysis for non-Gaussian states (i.e. those that cannot be completely discribed by their first and second statistical moments) and Gaussian states (the group we will be concerned in this work, where the first and second statistical moments completely describe the state, that in turn possess a Gaussian characteristic function, by definition) \cite{adesso2009optimal, Luca}. While in each studied case a specific quantum system can perform better or worse than other kind of state (for example, \cite{adesso2009optimal}), there is no ``rule'' dictating that some `special' quantum system is, in fact and in general, the golden state to perform all tasks better than classical ones.

As already mentioned \cite{aasi2013enhanced, dwyer2022squeezing, grote2013first}, the use of squeezed light in technological devices are of great importance nowadays. With this motivation as our driving force, in this work we focus our attention in the Quantum estimation problem within a Gaussian scenario: single- or two-mode states subjected to Gaussian dynamics. We are interested in the estimation of an unknown squeezing parameter encoded in one of the modes. We follow the strategy proposed by the authors in \cite{Luca}, where the mode to be encoded is sent to an Squeezer, and the other mode is an ancilla. The precision of estimating the squeezing parameter is quantified using the Average Quantum Fisher Information (AvQFI), which is computed by averaging over the phase acquired by the mode during its evolution. In \cite{Luca} the authors studied this problem both in complete generality and also for some classes of Gaussian states. Here we explicitly studied in details the most important classes of single- and two-mode Gaussian states as probes for this estimation proccess. For single-mode states we studied the interplay of the AvQFI and a \emph{Coherence} quantifier, as defined in \cite{xu2016quantifying}. Finally, for two-mode Gaussian states, we analliticaly and numerically investigate the AvQFI as a function of Entanglement.

This paper is organized as following: in section \ref{preliminares} we review in great details the formalism of Gaussian states, quantum estimation theory and the estimation strategy used in this work (following closely \cite{Luca}); in section \ref{results} we present our results, sorting by each class of Gaussian state we have chosen. Still in section \ref{results} we studied the AvQFI as a function of Coherence (for single-mode states), and as a function of the Logarithmic Negativity (for two-mode states). Finally, we conclude our work in section \ref{conclusions}.

\section{Preliminaries}\label{preliminares}

Our work is focused on estimating a parameter, specifically the squeezing parameter encoded in one mode of a Gaussian State. Therefore, in this section, we will review some important concepts concerning Gaussian States and Quantum Estimation Theory.

\subsection{Gaussian States}\label{gauss_sec}

In this work we are interested in single- or two-mode Gaussian States subjected to the so called Gaussian dynamics, i.e. dynamics that preserve the ``Gaussianity'' of the state (\cite{adesso2007entanglement}). A CV system of two modes $A$ and $B$ (with annihilation operators $\hat{a}$ and $\hat{b}$ respectively), can be defined by the quadrature vector $\hat{\boldsymbol{O}} = \{ \hat{q}_A, \hat{p}_A,  \hat{q}_B, \hat{p}_B\}$, where $\hat{q}_k = (\hat{a}_k + \hat{a}_k^\dagger)$ and $\hat{p}_k = i(\hat{a}_k^\dagger - \hat{a}_k)$, where $k = A, B$ (assuming natural units, $\hbar = 2$). We encourage the reader to pay attention to equations and definitions, since one can find both works with $\hbar =1$ and $\hbar = 2$ in the literature. The quadratures obey the canonical commutation relations $ [\hat{O}_j, \hat{O}_k] = 2i \Omega_{jk}$, with the two-mode symplectic form
\begin{equation}
\boldsymbol{\Omega} =
\left(
\begin{tabular}{cc}
0 & 1 \\                                                                                                                                                               -1 & 0 \\                                                                                                                                                              \end{tabular}                                                                                                                                                            \right)^{\bigoplus  2}.
\end{equation} The relations for single-mode Gaussian states are, obviouslly, the same, changing the dimension of the vectors and matrices.

A Gaussian state $\rho_{AB}$ \cite{adesso2007, Adesso2014ext, cerf2007, paris2005} is represented by a Gaussian characteristic function in phase space, and is completely characterized by its first and second statistical moments of the quadrature vector, given respectively by the displacement vector $\boldsymbol{\varepsilon}_{AB} = (\varepsilon_j)$ and the covariance matrix  $\boldsymbol{\sigma}_{AB} = (\sigma_{jk})$, where $\varepsilon_j = \langle \hat{O}_j \rangle$ and $\sigma_{jk} = \frac{1}{2} \langle \hat{O}_j \hat{O}_k + \hat{O}_k \hat{O}_j \rangle - \langle \hat{O}_j \rangle \langle \hat{O}_k \rangle$. A \emph{bona fide} condition satisfied by all physical Gaussian states is the Robertson-Schr\"{o}dinger uncertainty relation, given by
\begin{equation}\label{bonafide}
\boldsymbol{\sigma}_{AB} + i \boldsymbol \Omega \geq 0.
\end{equation}

For our purpose (since we will deal with mode B as an ancilla, as one can see soon in section \ref{strategy}), a general covariance matrix for a two-mode Gaussian state can be written as \cite{Luca}: \begin{equation}
\boldsymbol{\sigma}_{A B} = \left(                                                                                                                                                              \begin{tabular}{cccc}                                                                                                                                                               $a_1$ & $g$ & $c$ & 0 \\
$g$ & $b_1$ & 0 & $d$ \\
$c$ & 0 & $b_2$ & 0 \\
0 & $d$ & 0 & $b_2$
\end{tabular}                                                                                                                                                            \right),
\label{CMgeneral1}\end{equation} where the coeficients are defined such that (\ref{CMgeneral1}) satisfies the physical constraint (\ref{bonafide}). Its worth recalling that, by local symplectic operations (equivalent to local changes of basis on the state), every two-mode covariance matrix can be transformed to a standard form, with diagonal $2 \times 2$ subblocks that can be written as:
\begin{equation}
\boldsymbol{\sigma}_{A B} = \left(                                                                                                                                                              \begin{tabular}{cc}                                                                                                                                                               $\boldsymbol{\alpha}$ & $\boldsymbol{\gamma}$ \\                                                                                                                                                                $\boldsymbol{\gamma}^T$ & $\boldsymbol{\beta}$ \\                                                                                                                                                              \end{tabular}                                                                                                                                                            \right),
\label{CMgeneral}\end{equation}
where $\alpha = \textrm{diag} \{a,a\}$, $\beta = \textrm{diag} \{b,b\}$, $\gamma = \textrm{diag} \{c,d\}$, such that $a,b \geq 1$, $c \geq \abs{d} \geq 0.$ For future purposes, we define here the symplectic invariants: $A = \det \boldsymbol{\alpha}; B = \det \boldsymbol{\beta}; C = \det \boldsymbol{\gamma}; D = \det \boldsymbol{\sigma}_{AB}.$

The total mean number of excitations (proportional to the total mean energy) of a two-mode Gaussian state can be defined as: $E \equiv \bar{n}_A + \bar{n}_B = 2 \bar{n},$ where $\bar{n}_A = (\tr [\boldsymbol{\alpha}] - 2)/4 + (\varepsilon_{x,A}^2 + \varepsilon_{p,A}^2)/4$ and $\bar{n}_B = (\tr [\boldsymbol{\beta}] - 2)/4+(\varepsilon_{x,B}^2 + \varepsilon_{p,B}^2)/4$ are the mean number of excitations in modes $A$ and $B$ respectively (considering the displacement vector as non-null, since we will deal with such states soon), and $\bar{n}_i$ denotes the mean number of excitations per mode. Throughout this manuscript, we will consistently uphold the physical requirement that any initial state must be subject to a \emph{finite} mean energy constraint.

The various categories of Gaussian states that we address in this study will be detailed (including their covariance matrices, displacement vectors, etc.) in dedicated sections that follow.

To provide a comprehensive background on Gaussian states for the reader, ensuring a proper understanding of this work, we introduce two quantities that quantify important characteristics of Gaussian states. First, we will discuss an entanglement quantifier for Gaussian states, specifically the Logarithmic Negativity.

The \emph{Logarithmic Negativity}, $\mathcal{E}_N$, is a decreasing function of the smallest symplectic eigenvalue $\tilde{\nu}$ of the partial transpose of the covariance matrix: \be \mathcal{E}_N = \max \{ 0, - \ln \tilde{\nu} \}. \ee Here $\tilde{\nu}$ are the symplectic eigenvalues of the \emph{partially transposed covariance matrix} (for details we suggest the reader to check chapter 3 of \cite{adesso2007}), and one can show that the $\tilde{\nu}$ can be written in function of the symplectic invariants \cite{AdessoIP}: \be 2 \tilde{\nu}^2 = H - \sqrt{H^2 - 4D}\ee with $H = A + B - 2C$. The Logarithmic Negativity is a measure of Entanglement in composed Gaussian systems.

Finally, we introduce a coherence measure that we will use to analyze single-mode states. Coherence is typically associated with the concepts of interference and superposition states. However, how can we quantify a state's ability to formally exhibit \emph{Coherence}? This question was addressed a few years ago, in the discrete case by \cite{baumgratz2014quantifying} and for Gaussian states by \cite{xu2016quantifying}. Given a N-mode Gaussian state $\rho$, with covariance matrix $\boldsymbol{\sigma}$ and displacement vector $\boldsymbol{\varepsilon}$, the \emph{Coherence} of the state can be quantified by \cite{xu2016quantifying}: \be C(\rho) = - S(\rho) + \sum_{i=1}^N [(\bar{n_i} +1) \log_2 (\bar{n_i} +1) - \bar{n_i} \log_2 \bar{n_i}], \label{coherence_q}\ee where $\bar{n}_i$ is the mean photon number of the $i$-esim state, and $S(\rho)$ is the entropy of the state $\rho$: \be S(\rho) = - \sum_{i=1}^N \left[ \left( \frac{\nu_i - 1}{2} \right) \log_2 \left( \frac{\nu_i - 1}{2} \right) - \left( \frac{\nu_i + 1}{2} \right) \log_2 \left( \frac{\nu_i + 1}{2} \right) \right]. \label{entropy}\ee In equation (\ref{entropy}), $\{ \nu_i \}_{i=1}^N$ are the symplectic eigenvalues of the Covariance Matrix (please do not confuse $\nu$ with $\tilde{\nu}$, since the latter represents the symplectic eigenvalues of the partially transposed covariance matrix). In this framework, maximally coherent states are pure states. Furthermore, it is intuitive to observe that states exhibiting squeezing can have Coherence values (as defined in equation \ref{coherence_q}) greater than those of the so-called coherent states. As a final remark, we stress that we will differentiate \emph{Coherence} (the measure of Coherence as in Equation \ref{coherence_q}) from coherent (concerning coherent states) by capitalizing the first letter of the former.

\subsection{Quantum Estimation Theory}\label{qet}

For completeness, this subsection is dedicated to introducing Quantum Estimation Theory and the concept of Quantum Fisher Information (QFI). We encourage readers with a keen interest in exploring this topic further to consult the references such as \cite{helstrom1976quantum, holevo2003statistical, ballester2005estimation, paris2009quantum}.

The fundamental concept behind any estimation strategy, whether classical or quantum, is to obtain  information about a parameter encoded in the system with the highest possible accuracy. Typically, this information is acquired by sending N copies of the system to the encoding stage. In the final stage, the most precise measurement is performed on the system, aiming to obtain a value for the parameter that approaches the `actual' value. Consequently, an estimation strategy is grounded in statistical theory.

In this scenario, the idea is to encounter a probability distribution that depends on a parameter, denoted here as $p(x|\epsilon)$. The objective is to achieve the most accurate estimate of the parameter $\epsilon$ by repeatedly sampling the random variable $x$, which follows the distribution $p(x|\epsilon)$. An important principle in classical estimation theory, known as the \emph{Cramér-Rao bound}, asserts that the variance of any unbiased estimator $\hat{\epsilon}$ for the parameter $\epsilon$ must satisfy the following inequality:

\begin{equation}\label{classical_CRB}
\text{VAR}\left(\hat{\epsilon}\right) \geq \frac{1}{N F_\epsilon^M}.
\end{equation} where $N$ is the number of samplings, $\text{VAR}\left(\hat{\epsilon}\right)$ is the variance of the estimator $\hat{\epsilon}$ and $F_\epsilon^M$ is the classical Fisher information (where we explicitly showed the dependence of $F_\epsilon^M$ on the measurement M), defined by:
\begin{equation}
	F_\epsilon^M = \int \text{d}x \, p(x|\epsilon) \left[\partial_\epsilon \log p(x|\epsilon)\right]^2. 
\end{equation}

In a quantum version of the previously mentioned situation, the parameter $\epsilon$ is encoded in a quantum state $\rho_\epsilon$, typically by applying a quantum map $\Phi_\epsilon$ to a known input probe $\rho$: $\rho_\epsilon = \Phi_\epsilon[\rho].$ To acquire information about the system and, consequently, about $\epsilon$, a generic Positive Operator-Valued Measurement (POVM) must be executed on $\rho_\epsilon$. The maximum precision achievable within the bounds of quantum mechanics for unbiased estimation of $\epsilon$ is attained through optimization across all possible POVMs. 

This approach leads to the derivation of the quantum Cramér-Rao bound, which stipulates:
\begin{equation}\label{quantum_CRB}
\text{VAR}(\hat{\epsilon}) \geq \frac{1}{N H_\epsilon(\rho)}.
\end{equation} Here, $H_\epsilon$ represents the Quantum Fisher Information (QFI) linked to the encoded state $\rho_\epsilon$, derived from $\rho$. The Quantum Fisher Information is defined as:
\begin{equation}
H_\epsilon[\rho] = \Tr{\rho_\epsilon L_\epsilon^2}.
\end{equation} $L_\epsilon$ is the so called symmetric logarithmic derivative (SLD), an Hermitian operator that satisfies the relation:
\begin{equation}
\rho_\epsilon L_\epsilon + L_\epsilon \rho_\epsilon = 2 \,\partial_\epsilon \rho_\epsilon.
\end{equation} It can be showed \cite{ballester2005estimation, de2009estudo} that the Quantum Fisher Information (QFI) is linked to the second-order expansion of the Bures distance, or equivalently, the Uhlmann fidelity \\ ${\mathcal F (\rho_1,\rho_2)= \left(\Tr{\sqrt{\sqrt{\rho_1}\rho_2\sqrt{\rho_1}}}\right)^2}$ : \begin{equation}\label{QFI_Bures_expansion}
H_\epsilon[\rho] = 8 \lim_{\text{d}\epsilon\to 0} \frac{1-\sqrt{\mathcal F(\rho_\epsilon,\rho_{\epsilon+\text{d}\epsilon})}}{\text{d}\epsilon^2}.
\end{equation}

It's worth noticing two interesting aspects: (i) In general, Fisher Information is less than or equal to Quantum Fisher Information: $F_\epsilon^M \leq H_\epsilon$. (ii) By harnessing quantum resources such as entanglement, squeezing, ``genuine'' quantum states like Fock states, and others (as exemplified in \cite{adesso2009optimal, giovannetti2011advances, giovannetti2006quantum}), one can surpass the Standard Limit of estimation (equation \ref{classical_CRB}) (often referred to as the shot noise limit or standard quantum limit), and achieve the Heisenberg limit, i.e., reaching the saturation of equation \ref{quantum_CRB}. Moreover, employing quantum resources can lead to a quadratic improvement in estimation problems compared to using classical resources. While this quadratic gain is not a strict rule, it serves as a guiding principle in quantum system research aimed at parameter estimation.

We can attribute the following interpretation to Fisher information: it quantifies the sensitivity of a probability distribution (or a quantum state, in the quantum version) to small changes in the parameter $\epsilon$. If we make a small alteration $\theta$ to the probability distribution, and it becomes substantially different from the original distribution, this results in a higher Fisher information. Consequently, by the Cramér-Rao bound (and the Bures distance showed above), we can infer that we are moving away from the original probability distribution (or the original state), and therefore Fisher information quantifies the error with respect to the estimator: the higher the Fisher information, more the Fisher information will ``capture'' any deviation from the original distribution or state, the lower the variance with respect to the estimator.

\subsection{Estimation Strategy}\label{strategy}

\begin{figure}[h]
 \begin{center}
 \includegraphics[scale=0.35]{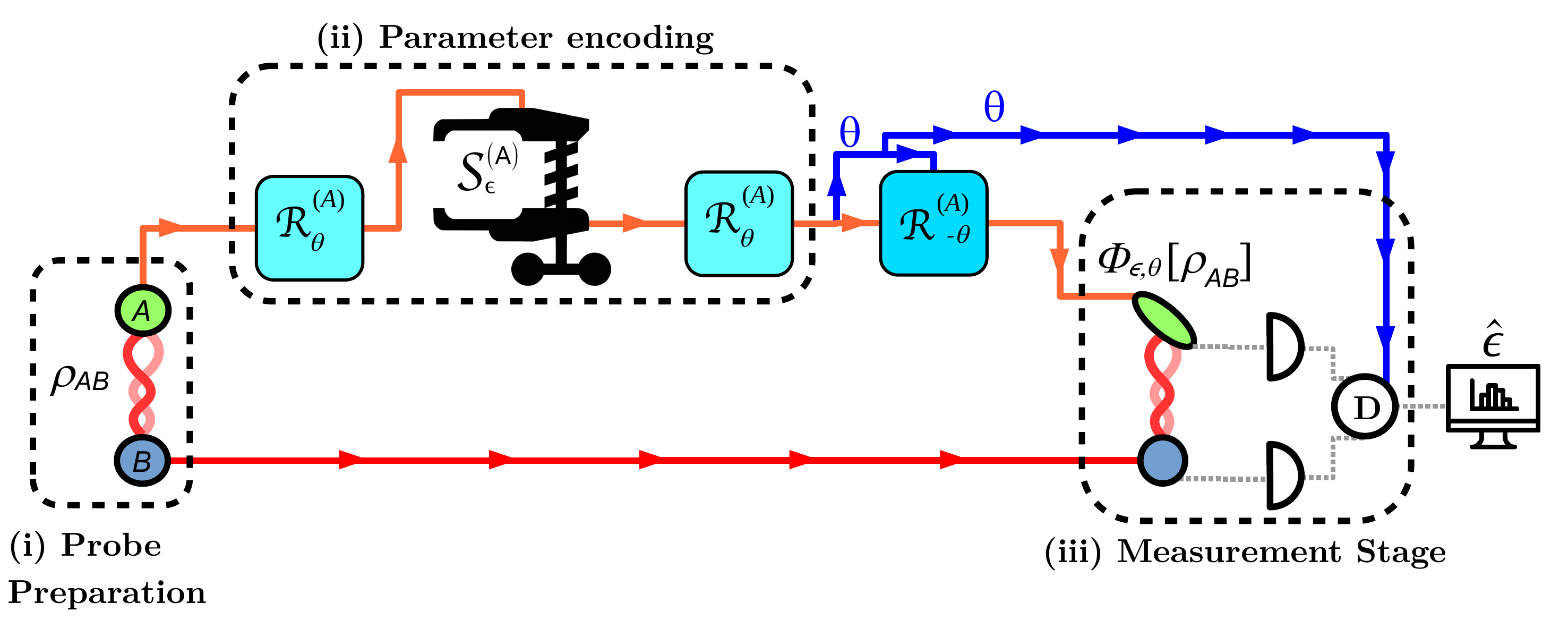}
 \end{center}
 \caption{(Color online)Proposed Estimation Strategy. The state is initially prepared as a single- or two-mode Gaussian state. Mode A is sent to a Squeezer, where the parameter $\epsilon$ is encoded. Mode B propagates freely. After the action of the Squeezer, mode A is brought back to be measured with mode B. Information about the phase $\theta$ acquired by the modes during their journey from state preparation to the measurement stage is utilized to establish the strategy. The average Quantum Fisher Information, $\overline{H_\epsilon}(\rho)$, is constructed by averaging over all phases $\theta$ in order to estimate the squeezing parameter $\epsilon$. More details are provided in section \ref{strategy} of the text.}\label{fig1}
\end{figure}

In this work we follow the estimation strategy as proposed in \cite{Luca}, depicted in Figure \ref{fig1}. Initially, the state is prepared as a single- or two-mode Gaussian state, $\rho_{AB}$, depending on the specific case under consideration. In this context, we designate mode A as the mode where the squeezing parameter will be encoded, and mode B as an ancilla. Naturally, mode B will not be considered when studying a single-mode Gaussian probe. In the next step, mode A is sent to a Squeezer, while mode B evolves freely. Just before the action of the Squeezer, the evolved state is given by: $
(\mathcal{R}_A \otimes \mathcal{R}_B) \rho_{AB} (\mathcal{R}_A \otimes \mathcal{R}_B)^\dagger,
$ where $\mathcal{R}_{A(B)} = e^{-i \theta_{A(B)}(t) a^\dagger a}$ is a unitary single-mode rotation operator acting on mode A(B), and acquired during the `flight' between the stages:

\be \mathcal{R}_\theta^{(A)} [\rho_A] = U_\theta^{(A)}[\rho_A] = e^{-i \theta a^\dagger a} \rho_A e^{i \theta a^\dagger a} =  \left( \begin{tabular}{cc} $\cos \theta$ & $\sin \theta$ \\ $-\sin \theta$ & $\cos \theta$  \end{tabular} \right).\nonumber \ee

After the first `flight', the Squeezer $\mathcal{S}_\epsilon^{(A)}=\exp[{\frac{\epsilon}{2}(a^2 - (a^\dagger)^2)}]$ acts on mode A, encoding the parameter $\epsilon$ that we are interested in estimate. Explicitly: \be \mathcal{S}_\epsilon^{(A)} [\rho_A] = U_\epsilon [\rho_A] = e^{-\frac{\epsilon}{2}(a^{\dagger 2}- a^2)} \rho_A e^{+\frac{\epsilon}{2}(a^{\dagger 2}- a^2)} =  \left( \begin{tabular}{cc} $e^\epsilon$ & $0$ \\ $0$ & $e^{-\epsilon}$ \end{tabular} \right).\nonumber \ee Mode A now returns, in order to be measured, acquiring another phase $\mathcal{R}_A$, and mode B a phase $\mathcal{R}_B$. The state can be written as: $$(\mathcal{R}_A \mathcal{S}_\epsilon^{(A)}\mathcal{R}_A \otimes \mathcal{R}_B^2) \rho_{AB}(\mathcal{R}_A \mathcal{S}_\epsilon^{(A)}\mathcal{R}_A \otimes \mathcal{R}_B^2)^\dagger.$$

If $\theta_{A,B}(t)$ is known by parts A and B, a unitary operation $(\mathcal{R}_A^{\dagger 2} \otimes \mathcal{R}B^{\dagger 2})$ can be applied by both parts. It is worth stressing this point of the proposed strategy: the knowledge of $\theta{A(B)}$ allows parts A and B to apply this unitary operation $(\mathcal{R}_A^{\dagger 2} \otimes \mathcal{R}_B^{\dagger 2})$ in such a way as to `remove' the dynamical phase introduced in mode B (the ancillary mode), leaving only the dynamics to which mode A is subjected. One can now write the state as: $$(\mathcal{R}_A^\dagger \mathcal{S}_\epsilon^{(A)}\mathcal{R}_A \otimes \mathds{1}_B) \rho_{AB}(\mathcal{R}_A^\dagger \mathcal{S}_\epsilon^{(A)}\mathcal{R}_A \otimes \mathds{1}_B)^\dagger.$$ The total dynamical map describing the full evolution of the state is: $\Phi_{\epsilon, \theta}[\rho_{AB}] = \Phi_{\epsilon, \theta}^A \otimes \mathds{1}^B [\rho_{AB}]$.

Since we are dealing with estimation theory \cite{paris2009quantum}, an estimator $\hat \epsilon$ for the applied squeezing can now be obtained (usually the maximum likelihood estimator is constructed), after a collection of measurements (if optimal, they can in principle minimize the error). The estimation accuracy of $\hat \epsilon$ is given by the quantum Cramer-Rao bound (as detailed in section \ref{qet}): \be \delta \hat \epsilon \geq \frac{1}{\sqrt{M H_\epsilon^{(\theta)}(\rho)}}\ee where $H_\epsilon^{(\theta)}(\rho)$ is the Quantum Fisher Information (QFI).

Depending on the phase $\theta$ acquired in flight, the estimation of $\epsilon$ can be different for each time-of-flight, therefore the \emph{versatility} of an input state can be checked by looking at the so called average QFI (AvQFI): \be \overline{H_\epsilon}(\rho) \equiv \int_0^{2 \pi} H_\epsilon^{(\theta)}(\rho) \frac{d \theta}{2 \pi}.\ee In \cite{Luca} the authors showed several general results concerning $\overline{H_\epsilon}(\rho)$, where we enphasize that by studying $\overline{H_\epsilon}(\rho)$ on can estimate the parameter $\epsilon$ within this approach: the average QFI $\overline{H_\epsilon}(\rho)$ sets a lower bound on the average value of $\delta \hat \epsilon$: \be \overline{\delta \hat \epsilon} \equiv \int_0^{2 \pi} \frac{d \theta}{2 \pi} \delta \hat \epsilon(\theta) \geq \frac{1}{\sqrt{M \overline{H_\epsilon}(\rho)}}\ee

In this paper we study the average QFI, $\overline{H_\epsilon}(\rho)$, for inportant classes of single and two-mode Gaussian States, within the estimation strategy proposed in this section.

\section{Results}\label{results}

In this section, we exploit our findings to present relevant analytical and numerical results. Our investigation covers both single-mode and two-mode states as a probe in our estimation strategy. For clarity, we have organized this section into subsections dedicated to each type of state: single-mode and two-mode states. Within each category, we provide a comprehensive overview of their general characteristics and present our results.

\subsection{Single-mode States}\label{single_mode}

\begin{figure}[!]
 \centering
 \includegraphics[scale=0.7]{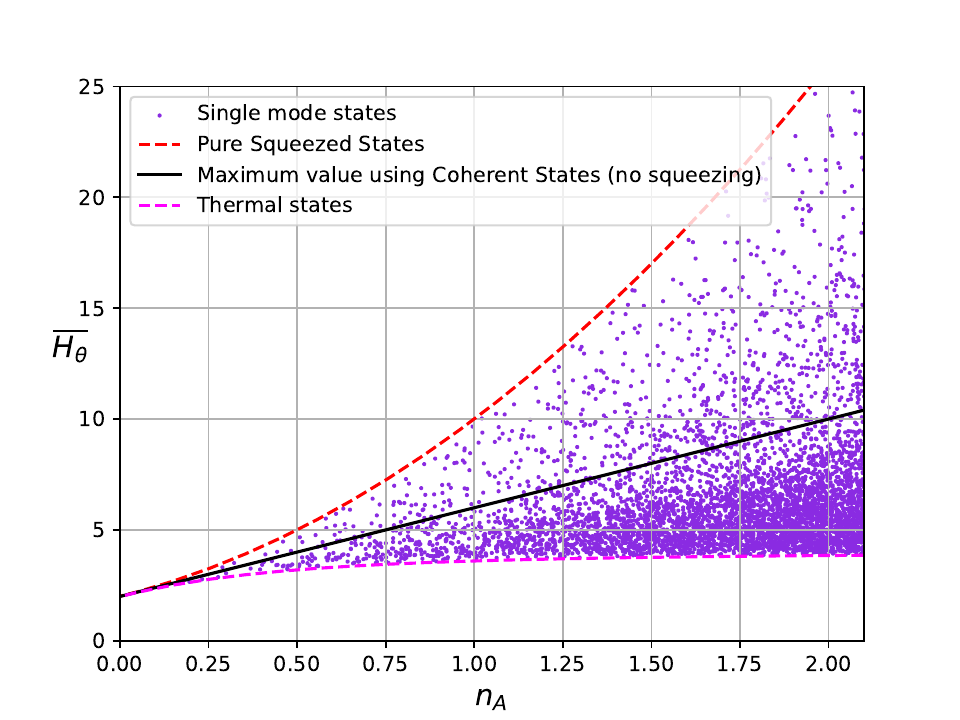}
 \caption{(Color online) Average QFI $\overline{H}_\theta$ in function of the mean photon number $n_A$ of the mode. Each point is a general single-mode Gaussian state, with parameters of the CM and displacement vector randomly chosen ($10^5$ states). The red dashed line is the upper bound for this estimation strategy, using $\overline{H}_\theta$, and is returned by pure squeezed states (see section \ref{single_mode}). The pink dashed curve is the lower bound, achieved by thermal states. The black solid curve is the maximum value of $\overline{H}_\theta$ using coherent states as probes (section \ref{single_coherent}).}
 \label{fig_single_general}
\end{figure}

A general single-mode Gaussian state is characterized by the following covariance matrix (CM): \begin{equation}
\boldsymbol{\sigma}_{A} = \left(
\begin{tabular}{cc}                                                                                                                                                               a & g \\
g & b 
\end{tabular}                                                                                                                                                            \right),
\label{CMgeneralsingle}\end{equation} and by the displacement vector: $\boldsymbol{\varepsilon} = (\varepsilon_x, \varepsilon_y).$ The parameters are chosen such that the CM represent a physical state, i.e. equation \ref{bonafide} is satisfied. If we consider, as probes, general single-mode Gaussian states, we obtain the results in Figure \ref{fig_single_general}. In this Figure we show the average QFI $\overline{H}_\theta$ in function of the mean photon number $n_A$ of the mode. Each point is a general single-mode Gaussian state, with parameters of the CM and displacement vector randomly chosen ($10^5$ states).

One can see that all states lie within the region bounded by the lower and upper limiting curves. The upper bound is given by:
\be \overline{H_\theta} = 4 n_A^2 + 4 n_A + 2,\label{upper_bound_single} \ee while the lower bound is given by: \be \overline{H_\theta} = 4 \frac{(2 n_A + 1)^2}{1 + (2 n_A + 1)^2}. \label{lower_bound_single} \ee It is interesting to highlight the physical significance of the upper bound: this curve is produced by single-mode pure squeezed states. Consequently, this class of states serves as the optimal probe for estimating the squeezing parameter $\varepsilon$ within the proposed strategy. One possible interpretation is that a pure squeezed state will be more sensitive to any parameter introduced into the state. The lower bound is produced by mixed thermal states.

\subsubsection{Coherent States}\label{single_coherent}

\begin{figure}[!]
 \centering
 \includegraphics[scale=0.7]{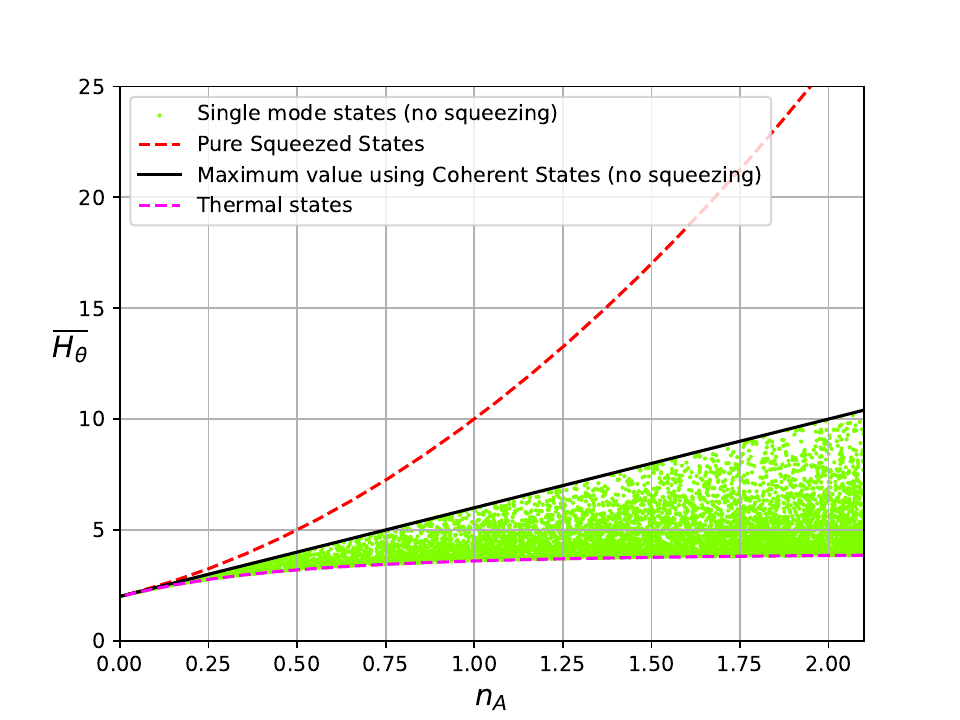}
 \caption{(Color online) Average QFI $\overline{H}_\theta$ in function of the mean photon number $n_A$ of the mode. Each point is a single-mode Gaussian coherent state, with parameters of the CM and displacement vector randomly chosen ($10^5$ states). The red dashed line is the upper bound for this estimation strategy, using $\overline{H}_\theta$, and is returned by pure squeezed states. The pink dashed curve is the lower bound, achieved by thermal states. The black solid curve is the maximum value of $\overline{H}_\theta$ using coherent states as probes (section \ref{single_coherent}).}
 \label{fig_single_coherent}
\end{figure}

Now we restrict our analysis to single-mode Gaussian coherent states as probe states, without squeezing, with covariance matrix: \begin{equation}
\boldsymbol{\sigma}_{A} = \left(
\begin{tabular}{cc}                                                                                                                                                               a & 0 \\
0 & a 
\end{tabular}                                                                                                                                                            \right),
\label{CMgeneralcoherent}\end{equation} and displacement vector $\boldsymbol{\varepsilon} = (\varepsilon_x, \varepsilon_y).$ Working out the same procedure with this class of states, we can see that the maximum value of $\overline{H}_\theta$ is given by the black solid curve in Figure \ref{fig_single_coherent}. Explicitly, this curve is given by: \be \overline{H_\theta} = 4 n_A + 2. \label{coherent_H}\ee

It is interesting that pure squeezed states approaches the Heisenberg limit \cite{Luca, escher2011general}, since $\overline{H}_\theta \sim n_A^2$, while using coherent states we can obtain $\overline{H}_\theta \sim n$, the so called Standard Quantum Limit (or Shot Noise Limit). This can be viewed as a ``quantum advantage'' in this metrology scheme.

\subsubsection{Relation between the Average QFI and Coherence of the state}

\begin{figure}[!]
 \centering
 \includegraphics[scale=0.5]{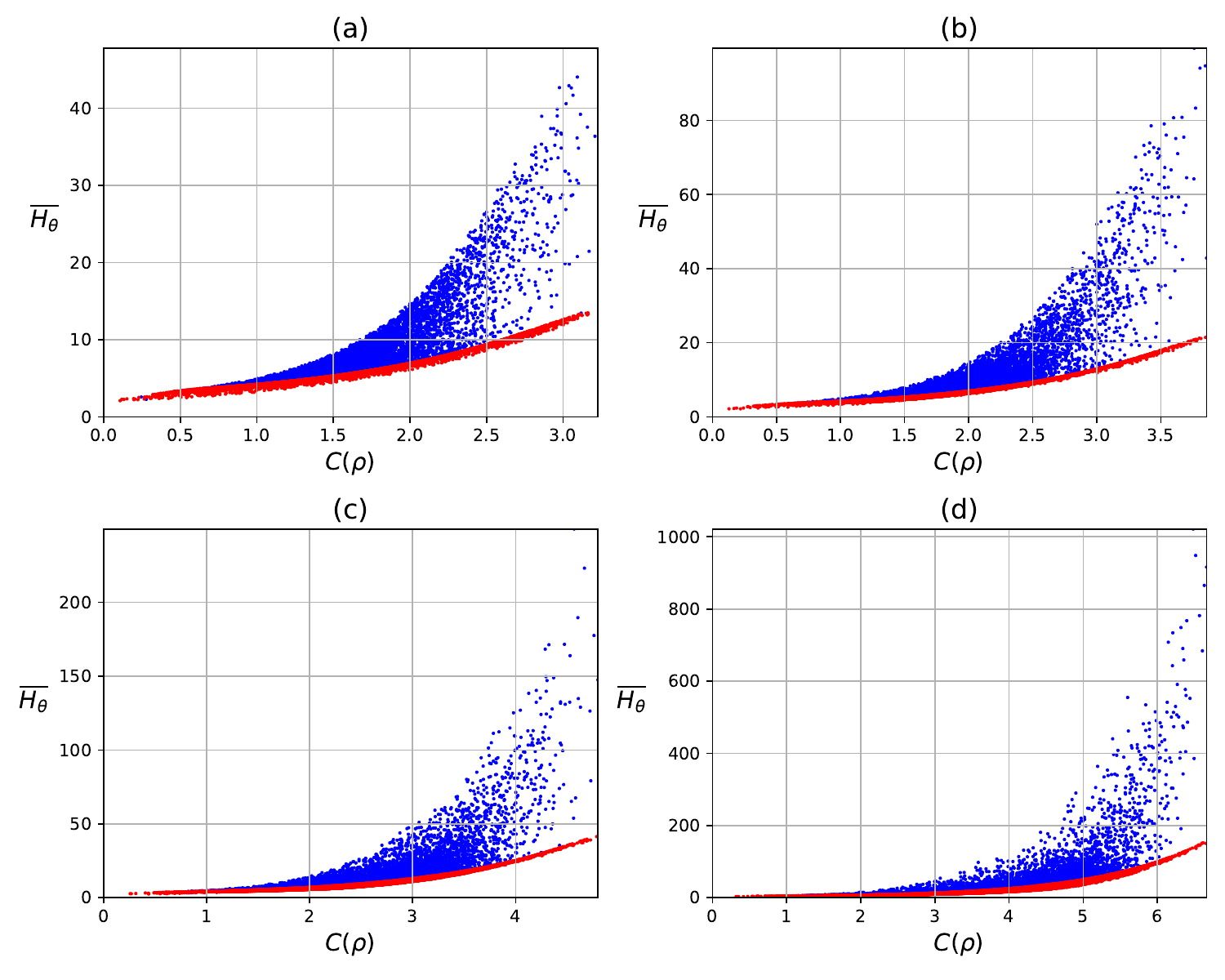}
 \caption{(Color online) Average Quantum Fisher Information ($\overline{H}_\theta$) as a function of Coherence ($C(\rho)$). The blue dots represent general single-mode states, while the red dots represent coherent single-mode states (a total of $10^5$ states were randomly chosen for each case). We varied the mean photon number ($n_A$) in each plot: (a) $n_A = 3$; (b) $n_A = 5$; (c) $n_A = 10$; (d) $n_A = 100$. It is evident that as $n_A$ increases, the value of $\overline{H}_\theta$ also increases. Furthermore, single-mode states with squeezing surpass coherent states in this estimation problem when they possess the same level of Coherence.}
 \label{fig_coherence_avqfi}
\end{figure}

A Coherence quantifier for Gaussian states was proposed in reference \cite{xu2016quantifying} and we briefly summarized in section \ref{gauss_sec}. The quantifier is given by equation \ref{coherence_q}: \be C(\rho) = - S(\rho) + \sum_{i=1}^N [(\bar{n_i} +1) \log_2 (\bar{n_i} +1) - \bar{n_i} \log_2 \bar{n_i}],\ee with $S(\rho)$ the relative entropy. Here we establish a relation between $\overline{H}_\theta$ and $C(\rho)$ for single-mode states. Our results are presented in Figure \ref{fig_coherence_avqfi}. 

In Figure \ref{fig_coherence_avqfi}, one can observe that coherent states achieve lower values of $\overline{H}_\theta$ compared to general single-mode states (which generally may exhibit some degree of squeezing) for the same amount of Coherence as defined in Equation \ref{coherence_q}. This corroborates the intuition that coherent states, being "quasi-classical" states, cannot attain the same level of sensitivity in this estimation problem when compared to squeezed states, which possess a ``quantum'' advantage.

\subsection{Two-mode States}

\begin{figure}[!]
 \centering
 \includegraphics[scale=0.7]{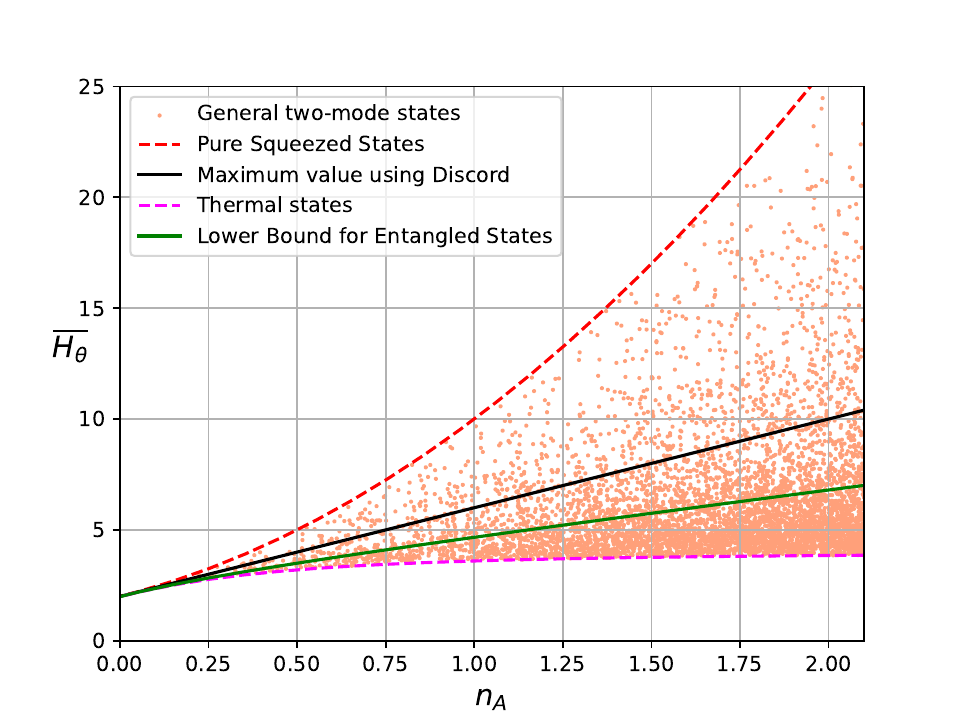}
 \caption{(Color online) Average QFI $\overline{H}_\theta$ in function of the mean photon number $n_A$ of mode A. Each point is a general two-mode Gaussian state, with parameters of the CM and displacement vector randomly chosen ($10^5$ states). The red dashed line is the upper bound for this estimation strategy, using $\overline{H}_\theta$, and is returned by pure two-mode squeezed states (see section \ref{sec_two_mode_squeezed}). The pink dashed curve is the lower bound, achieved by mixed thermal states. The black solid curve is the maximum value of $\overline{H}_\theta$ using states containing discord tipe correlations as probes (section \ref{sec_two_mode_discord}). Finally, the solid green curve is the best possible estimation value of $\overline{H}_\theta$ using states without any type of correlations (section \ref{sec_separable}).}
 \label{fig_general_two_mode}
\end{figure}

As mentioned before (equation \ref{CMgeneral1}), a general two-mode Gaussian state can be characterized by the following CM: \begin{equation}
\boldsymbol{\sigma}_{A B} = \left(                                                                                                                                                              \begin{tabular}{cccc}                                                                                                                                                               $a_1$ & $g$ & $c$ & 0 \\
$g$ & $b_1$ & 0 & d \\
$c$ & 0 & $b_2$ & 0 \\
0 & $d$ & 0 & $b_2$
\end{tabular}                                                                                                                                                            \right), \label{CMgeneral2}
\end{equation} with displacement vector given by: $\boldsymbol{\varepsilon} = (\varepsilon_x, \varepsilon_y, 0, 0).$ Tipically, general two-mode states has the same behavior of single-mode states. The upper and lower bounds are given by the same states, pure two-mode squeezed states and mixed thermal states, and by the exact same equations, Eq. \ref{upper_bound_single} and Eq. \ref{lower_bound_single}, respectively. Figure \ref{fig_general_two_mode} depicts how $10^5$ randomly chosen states are distributed, when we study $\overline{H}_\theta$ as a function of $n_A$. However, using two-mode states one can access more physics as we can see in the next sections.

\subsubsection{Separable States in the standard form}\label{sec_separable}

\begin{figure}[!]
 \centering
 \includegraphics[scale=0.7]{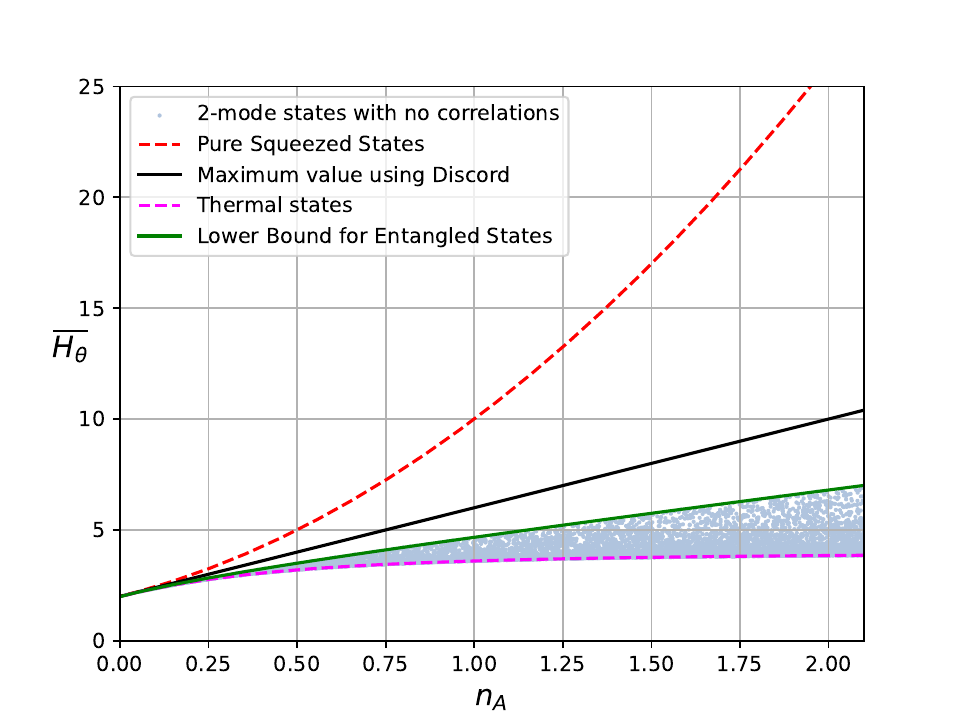}
 \caption{(Color online) Average QFI $\overline{H}_\theta$ in function of the mean photon number $n_A$ of mode A. Each point is an uncorrelated two-mode Gaussian state with CM given by equation \ref{separable_standard}, with parameters of the CM and displacement vector randomly chosen ($10^5$ states). The red dashed line is the upper bound for this estimation strategy, using $\overline{H}_\theta$, and is returned by pure two-mode squeezed states (see section \ref{sec_two_mode_squeezed}). The pink dashed curve is the lower bound, achieved by mixed thermal states. The black solid curve is the maximum value of $\overline{H}_\theta$ using states containing discord tipe correlations as probes (section \ref{sec_two_mode_discord}). Finally, the solid green curve is the best possible estimation value of $\overline{H}_\theta$ using states without any type of correlations (section \ref{sec_separable}).}
 \label{fig_separable_two_mode}
\end{figure}

We start with the simplest case: separable states with no correlations between modes A and B. Separable states in the standard form are characterized, within our approach, by: \begin{equation}
\boldsymbol{\sigma}_{A B} = \left(                                                                                                                                                              \begin{tabular}{cccc}                                                                                                                                                               $a_1$ & 0 & 0 & 0 \\
0 & $b_1$ & 0 & 0 \\
0 & 0 & $b_2$ & 0 \\
0 & 0 & 0 & $b_2$
\end{tabular}                                                                                                                                                            \right), \label{separable_standard}
\end{equation} with displacement vector given by: $\boldsymbol{\varepsilon} = (\varepsilon_x, \varepsilon_y, 0, 0).$ Our results for this class of states are depicted in Figure \ref{fig_separable_two_mode}. Interesting enough, for the case of separable states in the standard form as Equation \ref{separable_standard}, while obviously the lower bound is given by mixed thermal states (equation \ref{lower_bound_single}), the upper bound for separable states is given by (the solid green curve of Figure \ref{fig_separable_two_mode}): \be \overline{H_\theta} = 3 - \frac{1}{1 + 2 n_A} + 2 n_A. \label{upper_separable_standard} \ee This class of state can be thought as been worst than single-mode coherent states for this estimation problem. Naturally, if we allow our probe to be in the form of equation \ref{CMgeneral2}, with $c=d=0$, the state will behave as a single-mode coherent state and achieve the same result of equation \ref{coherent_H}. We can conjecture that, for this case of separable states in the standard form, some level of Coherence in mode A is required to reach greater level of squeezing estimation.

\subsubsection{Two-mode states with Discord type correlation}\label{sec_two_mode_discord}

\begin{figure}[!]
 \centering
 \includegraphics[scale=0.7]{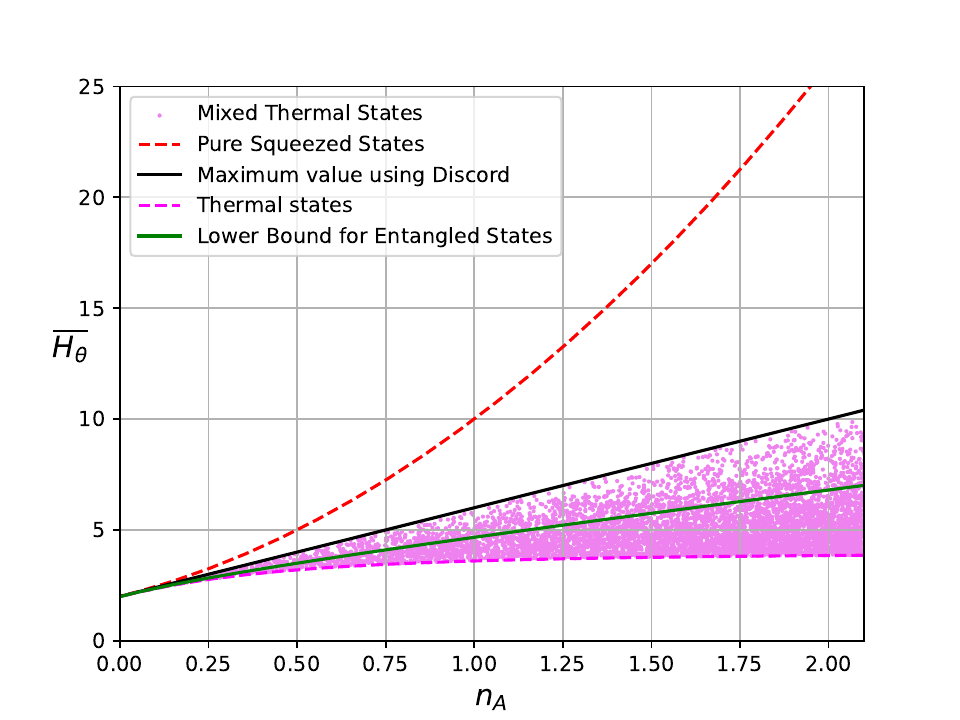}
 \caption{(Color online) Average QFI $\overline{H}_\theta$ in function of the mean photon number $n_A$ of mode A. Each point is a two-mode Gaussian state with CM given by equation \ref{discordant}, with parameters of the CM and displacement vector randomly chosen ($10^5$ states). The red dashed line is the upper bound for this estimation strategy, using $\overline{H}_\theta$, and is returned by pure two-mode squeezed states (see section \ref{sec_two_mode_squeezed}). The pink dashed curve is the lower bound, achieved by mixed thermal states. The black solid curve is the maximum value of $\overline{H}_\theta$ using states containing discord tipe correlations as probes (section \ref{sec_two_mode_discord}). Finally, the solid green curve is the best possible estimation value of $\overline{H}_\theta$ using states without any type of correlations (section \ref{sec_separable}).}
 \label{fig_discord_two_mode}
\end{figure}

Quantum Discord represents quantum correlations within composite systems that do not necessarily involve entanglement \cite{zurek2000einselection, ollivier2001quantum, henderson2001classical}. In the literature one can encounter a wide range of options to quantify and study Discord-type correlations. For example, it has been proven that, in CV systems, the so-called interferometric power is a measure of discord-type correlations for general mixed states $\rho_{AB}$, reducing to a measure of entanglement in the particular case of pure states \cite{AdessoIP, Bera2014}. A Discordant Gaussian state can be characterized by the following CM:
\begin{equation}
\boldsymbol{\sigma}_{A B} = \left(                                                                                                                                                              \begin{tabular}{cccc}                                                                                                                                                               a & 0 & c & 0 \\
0 & a & 0 & c \\
c & 0 & b & 0 \\
0 & c & 0 & b
\end{tabular}                                                                                                                                                            \right), \label{discordant}
\end{equation} with displacement vector given by: $\boldsymbol{\varepsilon} = (\varepsilon_x, \varepsilon_y, 0, 0),$ and $c \neq 0$. For this class of states our results are shown in Figure \ref{fig_discord_two_mode}. One can see clearly that this class of state reach the same maximum value for $\overline{H}_\theta$ as single-mode coherent states, given by equation \ref{coherent_H}.

\subsubsection{Entangled Two-mode States}\label{sec_two_mode_squeezed}

\begin{figure}[!]
 \centering
 \includegraphics[scale=0.7]{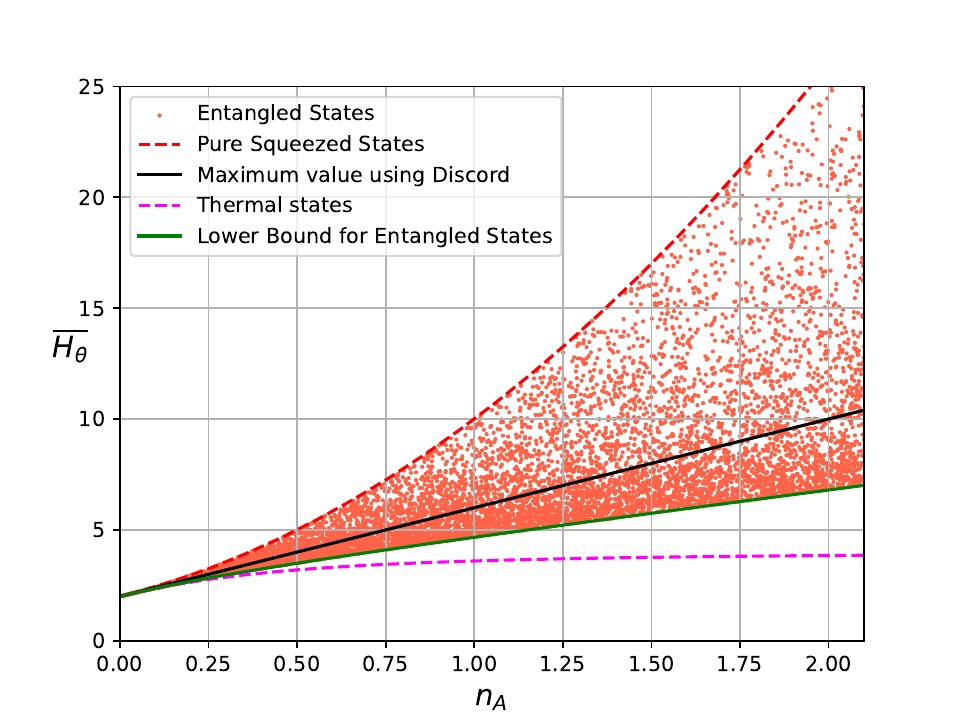}
 \caption{(Color online) Average QFI $\overline{H}_\theta$ in function of the mean photon number $n_A$ of mode A. Each point is an Entangled two-mode Gaussian state with CM given by equation \ref{CMEntangled}, with parameters of the CM and displacement vector randomly chosen ($10^5$ states). The red dashed line is the upper bound for this estimation strategy, using $\overline{H}_\theta$, and is returned by pure two-mode squeezed states (see section \ref{sec_two_mode_squeezed}). The pink dashed curve is the lower bound, achieved by mixed thermal states. The black solid curve is the maximum value of $\overline{H}_\theta$ using states containing discord tipe correlations as probes (section \ref{sec_two_mode_discord}). Finally, the solid green curve is the best possible estimation value of $\overline{H}_\theta$ using states without any type of correlations (section \ref{sec_separable}).}
 \label{fig_entangled}
\end{figure}

In this section, we examine the last category of Gaussian states we have investigated in this study. We dedicate our attention now to the important class of Entangled states in the standard form, where the CM is given by: \begin{equation}
\boldsymbol{\sigma}_{A B} = \left(                                                                                                                                                              \begin{tabular}{cccc}                                                                                                                                                               a & 0 & c & 0 \\
0 & a & 0 & -c \\
c & 0 & b & 0 \\
0 & -c & 0 & b
\end{tabular}                                                                                                                                                            \right), \label{CMEntangled}
\end{equation} with displacement vector given by: $\boldsymbol{\varepsilon} = (0, 0, 0, 0).$ For this important class, that include the states that saturate our measure $\overline{H}_\theta$ (pure two-mode-squeezed states), our results show the same upper bound as single-mode Gaussian states (equation \ref{upper_bound_single}): \be \overline{H_\theta} = 4 n_A^2 + 4 n_A + 2, \ee while the lower bound is, in agreement with previous sections, the upper bound for separable states in the standard form (equation \ref{upper_separable_standard}). Figure \ref{fig_entangled} depicts our results for $\overline{H}_\theta$ in terms of $n_A$.

\subsubsection{Relation between the Average QFI and Entanglement}\label{sec_logneg}

\begin{figure}[!]
 \centering
 \includegraphics[scale=0.5]{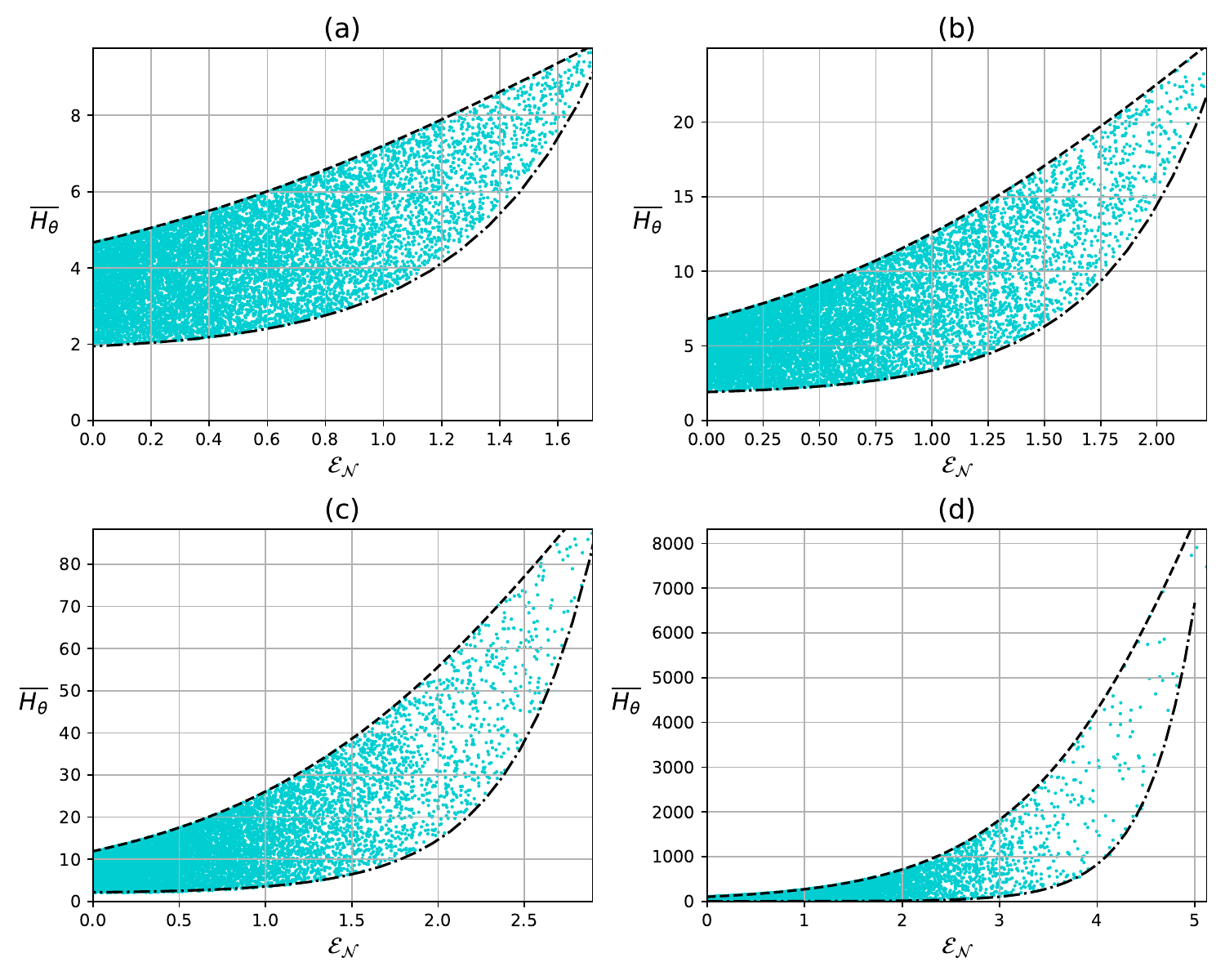}
 \caption{(Color online) Average Quantum Fisher Information ($\overline{H}\theta$) as a function of the Logarithmic Negativity ($\mathcal{E}_N$). The dots represent Entangled two-mode states (a total of $10^5$ states were randomly chosen for each case). We varied the mean photon number ($n_A$) in each plot: (a) $n_A = 3$; (b) $n_A = 5$; (c) $n_A = 10$; (d) $n_A = 100$. It is evident that as $n_A$ increases, the value of $\overline{H}_\theta$ also increases. The black dashed curves are the upper bound for each case, returned by pure squeezed two-mode states. The dotdashed black curves are the lower bound for each energy, and were obtained numerically. Details about the upper and lower bounds in the text.}
 \label{logneg}
\end{figure}

In order to obtain some physical intuition concerning the interplay between Entanglement and this estimation problem, we studied the relation between the Average QFI, $\overline{H}_\theta$, and the Entanglement quantifier mentioned in section \ref{gauss_sec}, explicitly the Logarithmic Negativity $\mathcal{E}_N$ \cite{AdessoIP}. We first note that our result is Energy dependent, i.e., we were able to obtain results that depend explicitly on the energy of mode A $n_A$. We focused our study in states with CM as in equation \ref{CMEntangled}. In Figure \ref{logneg} we show our results for different values of $n_A$. The overall value of $\overline{H}_\theta$ increase with $n_A$, an intuitive result, since if one use a higher value of Energy, more one can access the paramenter to be estimated.

In Figure \ref{logneg}, it is evident that as the system's Entanglement, measured by $\mathcal{E}_N$, increases, so does the value of $\overline{H}_\theta$. An important result of our work is that the upper bound for Figures \ref{logneg} is given by pure two-mode squeezed states, and we were able to obtain an analytical expression for this dependence as: \be \overline{H_\theta} = 2 + \frac{8n_A(1 + n_A)}{1 + (2 + 4n_A - \tilde{\nu}) \tilde{\nu}} =  2 + \frac{8n_A(1 + n_A)}{1 + (2 + 4n_A - \ei^{-\mathcal{E}_\mathcal{N}}) \ei^{-\mathcal{E}_\mathcal{N}}} \label{logneq_equation}, \ee where it is clear the dependence of $\overline{H}_\theta$ on the Entanglement measure $\mathcal{E}_N$ and also on the Energy of mode A, $n_A$.

To investigate the lower bound, numerical methods were necessary, as it is not straightforward to determine which states yield the lower bounds. For each dataset comprising $(\mathcal{E}_N, \overline{H}_\theta)$, we conducted a numerical regression analysis to model the lower bound, aiming to obtain a curve that approximates this bound. After conducting this analysis, we derived the following expression for the lower bound:
\be \overline{H}_\theta (\mathcal{E}_N) = A_1 \exp \left[ B_1 \mathcal{E}_N \right] + A_2.\ee

The specific values of $A_1$, $A_2$, and $B_1$ depend on each individual case, corresponding to different energy levels $n_A$. In Table \ref{table1}, we present the numerical results for each $n_A$ displayed in Figure \ref{logneg}. Additionally, we provide the Mean Square Error (MSE) and the standard deviation of MSE (which we call in this work $\Delta_{MSE}$). It is noteworthy that the lower bound for $n_A=10$ and $n_A=100$ are the same due to the characteristics of our numerical approach.

\begin{table}[h]
\caption{Numerical analysis of the lower bound of $\overline{H}_\theta$ in function of $\mathcal{E}_N$}
\begin{tabular}{d|d|d|d|d|d}
n_A & A_1 & A_2 & B_1 & MSE & \Delta_{MSE} \\
3 & 0.169 & 1.778 & 2.187 & 0.002 & 0.486 \\
5 & 0.215 & 1.695 & 2.037 & 0.007 & 0.586\\
10 & 0.193 & 1.964 & 2.089 & 0.402 & 0.963 \\
100 & 0.193 & 1.964 & 2.089 & 0.402 & 0.963 \\
\end{tabular}
\label{table1}
\end{table}

It can be observed that the MSE increases significantly as the energy values rise, while for low energy values, it remains close to zero. This indicates that our results are quite accurate for lower energy levels when analyzing the lower bound. Conversely, for higher values of $n_A$, both the MSE and $\Delta_{MSE}$ increase, and we understand that our analysis is enough for the context of this work. Finally we mention that it would be interest for the community if the specific class of states that return the lower bound could be obtained, in the same sense that pure two-mode squeezed states return the upper bound.

\section{Conclusion}\label{conclusions}

In this work we exploit important classes of single- and two-mode Gaussian states, for the specific problem of estimate with the higher precision the squeezing parameter $\epsilon$ in one of the modes. After quite complete sections reviewing Gaussian states, Estimation theory we detailed presented the estimation strategy studied in this work, within the approach as originally proposed by \cite{Luca} and followed closely here.

For single-mode states we showed that pure squeezed states are the best probes for this estimation problem, approaching the so called Heisenberg limit. Coherent states, been a ``quasi-classical'' class of states, achieve the linear (in relation to the Energy) behavior of the Shot Noise limit (or Quantum Standard limit). Still for single-mode states, we studied a relation between the state's Coherence and the Average QFI, showing that even among states with similar levels of coherence, those with squeezing capabilities can outperform coherent states in terms of estimation precision.

For two-mode Gaussian states, our results demonstrate the significance of entanglement in squeezing estimation. Pure two-mode squeezed states have the potential to approach the Heisenberg limit ($\overline{H}_\theta \sim n^2$), while states exhibiting Discord-type correlations yield $\overline{H}_\theta \sim n$. We have also explored other classes of state, including separable states in their standard form, and provided analytical results for upper and lower bounds in each category (in the analysis of $\overline{H}_\theta$ as a function of $n_A$). Furthermore, we conducted an investigation into the relationship between entanglement and the Average QFI, revealing intriguing findings: (i) As energy levels increase, the Average QFI also rises, enhancing the precision of the squeezing estimation problem. (ii) Once again, pure two-mode squeezed states emerge as the optimal probes for squeezing estimation, constituting the upper bound for states concerning the interplay between the Average QFI and Logarithmic Negativity. We also presented an analytical result for this upper bound. (iii) Numerical results for the lower bound within the study of the Average QFI and Entanglement (measured by $\mathcal{E}_N$) were presented.

We encourage the community to investigate whether there are classes of Gaussian states that meet this lower bound, similar to how pure two-mode squeezed states satisfy the upper bound.

\section*{Acknowledgments}
L.A.M.S. would like to express gratitude to INCT-IQ (Instituto Nacional de Ci\^encia e Tecnologia - Informa\c{c}\~ao Qu\^antica) for their financial support during the VIII Paraty Quantum Information School and Workshop. Special thanks are extended to Ana Mizher, Rodrigo Dias, and Guilherme Dinnebier for their invaluable contributions through insightful discussions on data analysis. Additionally, L.A.M.S. extends appreciation to Carlos Henrique S. Vieira and Irismar G. da Paz for their helpful insights and discussions regarding Coherence quantifiers. This work makes use of the QuGIT toolbox \cite{QuGIT}.






\bibliography{avqfi_v0}

\begin{thebibliography}{10}
\expandafter\ifx\csname url\endcsname\relax
  \def\url#1{\texttt{#1}}\fi
\expandafter\ifx\csname urlprefix\endcsname\relax\def\urlprefix{URL }\fi
\expandafter\ifx\csname href\endcsname\relax
  \def\href#1#2{#2} \def\path#1{#1}\fi

\bibitem{kubacek2012foundations}
L.~Kub{\'a}cek, Foundations of estimation theory, Elsevier, 2012.

\bibitem{kay1993fundamentals}
S.~M. Kay, Fundamentals of statistical signal processing: estimation theory,
  Prentice-Hall, Inc., 1993.

\bibitem{helstrom1976quantum}
C.~Helstrom, Quantum detection and estimation theory, Mathematics in Science
  and Engineering. New York: Academic Press 123 (1976).

\bibitem{holevo2003statistical}
A.~S. Holevo, Statistical structure of quantum theory, Vol.~67, Springer
  Science \& Business Media, 2003.

\bibitem{ballester2005estimation}
M.~A. Ballester~S{\'a}nchez, Estimation of quantum states and operations, Ph.D.
  thesis, Utrecht University (2005).

\bibitem{paris2009quantum}
M.~G. Paris, Quantum estimation for quantum technology, International Journal
  of Quantum Information 7~(supp01) (2009) 125--137.

\bibitem{adesso2009optimal}
G.~Adesso, F.~Dell’Anno, S.~De~Siena, F.~Illuminati, L.~Souza, Optimal
  estimation of losses at the ultimate quantum limit with non-gaussian states,
  Physical Review A 79~(4) (2009) 040305.

\bibitem{AdessoIP}
G.~Adesso, \href{https://link.aps.org/doi/10.1103/PhysRevA.90.022321}{Gaussian
  interferometric power}, Phys. Rev. A 90 (2014) 022321.
\newblock \href {https://doi.org/10.1103/PhysRevA.90.022321}
  {\path{doi:10.1103/PhysRevA.90.022321}}.
\newline\urlprefix\url{https://link.aps.org/doi/10.1103/PhysRevA.90.022321}

\bibitem{Luca}
L.~Rigovacca, A.~Farace, L.~A.~M. Souza, A.~De~Pasquale, V.~Giovannetti,
  G.~Adesso,
  \href{https://link.aps.org/doi/10.1103/PhysRevA.95.052331}{Versatile gaussian
  probes for squeezing estimation}, Phys. Rev. A 95 (2017) 052331.
\newblock \href {https://doi.org/10.1103/PhysRevA.95.052331}
  {\path{doi:10.1103/PhysRevA.95.052331}}.
\newline\urlprefix\url{https://link.aps.org/doi/10.1103/PhysRevA.95.052331}

\bibitem{luo2022environmental}
M.~Luo, W.~Liu, Y.~Chen, S.~Han, S.~Gao, Environmental parameter estimation
  with the two-level atom probes, Chinese Physics B 31~(5) (2022) 050304.

\bibitem{predko2020time}
A.~Predko, F.~Albarelli, A.~Serafini, Time-local optimal control for parameter
  estimation in the gaussian regime, Physics Letters A 384~(13) (2020) 126268.

\bibitem{assad2020accessible}
S.~M. Assad, J.~Li, Y.~Liu, N.~Zhao, W.~Zhao, P.~K. Lam, Z.~Ou, X.~Li,
  Accessible precisions for estimating two conjugate parameters using gaussian
  probes, Physical Review Research 2~(2) (2020) 023182.

\bibitem{bradshaw2018ultimate}
M.~Bradshaw, P.~K. Lam, S.~M. Assad, Ultimate precision of joint quadrature
  parameter estimation with a gaussian probe, Physical Review A 97~(1) (2018)
  012106.

\bibitem{morelli2021bayesian}
S.~Morelli, A.~Usui, E.~Agudelo, N.~Friis, Bayesian parameter estimation using
  gaussian states and measurements, Quantum Science and Technology 6~(2) (2021)
  025018.

\bibitem{ge2020operational}
W.~Ge, K.~Jacobs, S.~Asiri, M.~Foss-Feig, M.~S. Zubairy, Operational resource
  theory of nonclassicality via quantum metrology, Physical Review Research
  2~(2) (2020) 023400.

\bibitem{escher2011general}
B.~Escher, R.~L. de~Matos~Filho, L.~Davidovich, General framework for
  estimating the ultimate precision limit in noisy quantum-enhanced metrology,
  Nature Physics 7~(5) (2011) 406--411.

\bibitem{escher2011quantum}
B.~Escher, R.~de~Matos~Filho, L.~Davidovich, Quantum metrology for noisy
  systems, Brazilian Journal of Physics 41~(4-6) (2011) 229--247.

\bibitem{maleki2023speed}
Y.~Maleki, B.~Ahansaz, A.~Maleki, Speed limit of quantum metrology, Scientific
  Reports 13~(1) (2023) 12031.

\bibitem{schnabel2010quantum}
R.~Schnabel, N.~Mavalvala, D.~E. McClelland, P.~K. Lam, Quantum metrology for
  gravitational wave astronomy, Nature communications 1~(1) (2010) 121.

\bibitem{demkowicz2013fundamental}
R.~Demkowicz-Dobrza{\'n}ski, K.~Banaszek, R.~Schnabel, Fundamental quantum
  interferometry bound for the squeezed-light-enhanced gravitational wave
  detector geo 600, Physical Review A 88~(4) (2013) 041802.

\bibitem{pezze2018quantum}
L.~Pezze, A.~Smerzi, M.~K. Oberthaler, R.~Schmied, P.~Treutlein, Quantum
  metrology with nonclassical states of atomic ensembles, Reviews of Modern
  Physics 90~(3) (2018) 035005.

\bibitem{giovannetti2011advances}
V.~Giovannetti, S.~Lloyd, L.~Maccone, Advances in quantum metrology, Nature
  photonics 5~(4) (2011) 222--229.

\bibitem{giovannetti2006quantum}
V.~Giovannetti, S.~Lloyd, L.~Maccone, Quantum metrology, Physical review
  letters 96~(1) (2006) 010401.

\bibitem{abbott2020prospects}
B.~P. Abbott, R.~Abbott, T.~Abbott, S.~Abraham, F.~Acernese, K.~Ackley,
  C.~Adams, V.~Adya, C.~Affeldt, M.~Agathos, et~al., Prospects for observing
  and localizing gravitational-wave transients with advanced ligo, advanced
  virgo and kagra, Living reviews in relativity 23 (2020) 1--69.

\bibitem{yu2020quantum}
H.~Yu, L.~McCuller, M.~Tse, N.~Kijbunchoo, L.~Barsotti, N.~Mavalvala, Quantum
  correlations between light and the kilogram-mass mirrors of ligo, Nature
  583~(7814) (2020) 43--47.

\bibitem{aasi2013enhanced}
J.~Aasi, J.~Abadie, B.~Abbott, R.~Abbott, T.~Abbott, M.~Abernathy, C.~Adams,
  T.~Adams, P.~Addesso, R.~Adhikari, et~al., Enhanced sensitivity of the ligo
  gravitational wave detector by using squeezed states of light, Nature
  Photonics 7~(8) (2013) 613--619.

\bibitem{dwyer2022squeezing}
S.~E. Dwyer, G.~L. Mansell, L.~McCuller, Squeezing in gravitational wave
  detectors, Galaxies 10~(2) (2022) 46.

\bibitem{grote2013first}
H.~Grote, K.~Danzmann, K.~Dooley, R.~Schnabel, J.~Slutsky, H.~Vahlbruch, First
  long-term application of squeezed states of light in a gravitational-wave
  observatory, Physical review letters 110~(18) (2013) 181101.

\bibitem{niezgoda2021many}
A.~Niezgoda, J.~Chwede{\'n}czuk, Many-body nonlocality as a resource for
  quantum-enhanced metrology, Physical Review Letters 126~(21) (2021) 210506.

\bibitem{marciniak2022optimal}
C.~D. Marciniak, T.~Feldker, I.~Pogorelov, R.~Kaubruegger, D.~V. Vasilyev,
  R.~van Bijnen, P.~Schindler, P.~Zoller, R.~Blatt, T.~Monz, Optimal metrology
  with programmable quantum sensors, Nature 603~(7902) (2022) 604--609.

\bibitem{degen2017quantum}
C.~L. Degen, F.~Reinhard, P.~Cappellaro, Quantum sensing, Reviews of modern
  physics 89~(3) (2017) 035002.

\bibitem{xu2016quantifying}
J.~Xu, Quantifying coherence of gaussian states, Physical Review A 93~(3)
  (2016) 032111.

\bibitem{adesso2007entanglement}
G.~Adesso, Entanglement of gaussian states (2007).
\newblock \href {http://arxiv.org/abs/quant-ph/0702069}
  {\path{arXiv:quant-ph/0702069}}.

\bibitem{adesso2007}
G.~Adesso, F.~Illuminati, Entanglement in continuous-variable systems: recent
  advances and current perspectives, Journal of Physics A: Mathematical and
  Theoretical 40 (2007) 7821.
\newblock \href {https://doi.org/10.1088/1751-8113/40/28/S01}
  {\path{doi:10.1088/1751-8113/40/28/S01}}.

\bibitem{Adesso2014ext}
G.~Adesso, S.~Ragy, A.~R. Lee, Continuous variable quantum information:
  Gaussian states and beyond, Open Systems \& Information Dynamics 21 (2014)
  1440001.
\newblock \href {https://doi.org/10.1142/S1230161214400010}
  {\path{doi:10.1142/S1230161214400010}}.

\bibitem{cerf2007}
N.~J. Cerf, G.~Leuchs, E.~S. Polzik, Quantum Information with Continuous
  Variables of Atoms and Light, Imperial College Press, 2007.

\bibitem{paris2005}
A.~Ferraro, S.~Olivares, M.~G.~A. Paris, Gaussian states in continuous variable
  quantum information, Napoli Series on Physics and Astrophysics (ed.
  Bibliopolis, Napoli, 2005), 2005.

\bibitem{baumgratz2014quantifying}
T.~Baumgratz, M.~Cramer, M.~B. Plenio, Quantifying coherence, Physical review
  letters 113~(14) (2014) 140401.

\bibitem{de2009estudo}
L.~A.~M. de~Souza, Estudo de estados de vari{\'a}veis cont{\'\i}nuas gaussianos
  e n{\~a}o-gaussianos monomodais sob efeito de um canal dissipativo, Phd
  thesis, Universidade Federal de Minas Gerais (2009).

\bibitem{zurek2000einselection}
W.~H. Zurek, Einselection and decoherence from an information theory
  perspective, Annalen der Physik 512~(11-12) (2000) 855--864.

\bibitem{ollivier2001quantum}
H.~Ollivier, W.~H. Zurek, Quantum discord: a measure of the quantumness of
  correlations, Physical review letters 88~(1) (2001) 017901.

\bibitem{henderson2001classical}
L.~Henderson, V.~Vedral, Classical, quantum and total correlations, Journal of
  physics A: mathematical and general 34~(35) (2001) 6899.

\bibitem{Bera2014}
M.~N. Bera, Quantum fisher information as the measure of gaussian quantum
  correlation: Role in quantum metrology, arXiv preprint arXiv:1406.5144
  (2014).

\bibitem{QuGIT}
I.~Brandão, D.~Tandeitnik, T.~Guerreiro,
  \href{https://www.sciencedirect.com/science/article/pii/S0010465522001904}{Qugit:
  A numerical toolbox for gaussian quantum states}, Computer Physics
  Communications 280 (2022) 108471.
\newblock \href {https://doi.org/https://doi.org/10.1016/j.cpc.2022.108471}
  {\path{doi:https://doi.org/10.1016/j.cpc.2022.108471}}.
\newline\urlprefix\url{https://www.sciencedirect.com/science/article/pii/S0010465522001904}

\end{thebibliography}


\end{document}